\documentclass[aps,prl,twocolumn,showpacs]{revtex4}
\usepackage{graphicx}
\usepackage{multirow}
\usepackage{amsfonts}
\usepackage{amssymb}
\usepackage{amsmath}

\begin{document}
\title{Charge Order Breaks Magnetic Symmetry in Molecular Quantum Spin Chains}
\author{M.~Dressel}
\author{M.~Dumm}
\author{T.~Knoblauch}
\author{B.~K\"ohler}
\author{B.~Salameh}
\author{S.~Yasin}
\affiliation{1. Physikalisches Institut, Universit{\"a}t Stuttgart, ~Pfaffenwaldring 57, 70550 Stuttgart, Germany}

\date{\today}

\begin{abstract}
Charge order affects most of the electronic properties but is
believed not to alter the spin arrangement since the
magnetic susceptibility remains unchanged. We present
electron-spin-resonance
experiments on quasi-one-dimensional (TMTTF)$_2X$ salts ($X$=
PF$_6$, AsF$_6$ and SbF$_6$), which reveal that the magnetic
properties are modified below $T_{\rm CO}$ when electronic ferroelectricity sets in. The coupling of
anions and organic molecules rotates the ${\bf g}$-tensor out of
the molecular plane creating magnetically  non-equivalent sites on
neighboring chains at domain walls. Due to anisotropic Zeeman interaction a novel
magnetic interaction mechanism in the charge-ordered state is
observed as a doubling of the rotational periodicity of $\Delta H$.
\end{abstract}

\maketitle

\section{Introduction}
Charge disproportionation is a common ordering mechanism in
correlated electron systems that is intensively investigated in
transition-metal oxides as well as organic compounds
\cite{Imada98,Seo04} because it is recognized to influence charge,
lattice and magnetic degrees of freedom. For static charge order
(CO) the electrons are (partially) localized resulting in an
insulating state, most evident in manganites or at the $1/8$
anomaly observed in cuprates \cite{Tranquada95}. Fluctuating
stripes, however, may be relevant for high-temperature
superconductivity \cite{Kivelson03}, but also in organic systems
evidence accumulates that charge fluctuations  mediate
superconductivity \cite{Merino01}.

As far as the magnetic properties are concerned, the situation is
even more puzzling. CO can induce ferroelectricity when the
inversion symmetry is broken, as suggested (and also
disputed) for  some mixed-valence
oxides \cite{Ikeda05,Niermann12}, magnetite \cite{Alexe09,Yamauchi09,Schrettle11},
manganites \cite{Efremov04,Choi08} or
nickelates \cite{Ramesh07,Cheong07}. For perovskites, however, there seems
to be a mutual exclusion of magnetism and ferroelectricity
\cite{Hill00,Khomskii01}, albeit the field of multiferroics drew
enormous attention over the last years \cite{Kimura03,Loidl08}. About ten
years ago, electronic ferroelectricity was discovered in
quasi-one-dimensional organic compounds (TMTTF)$_2X$ (where TMTTF
stands for tetra\-methyl\-tetra\-thia\-fulvalene and $X$ denotes a
monovalent anion such as PF$_6$, AsF$_6$, SbF$_6$, Br) around
$50-160$~K \cite{Monceau01}. X-ray investigations could not
identify any superstructure, on this account it is known as a
``structureless'' transition \cite{Coulon85}.

\begin{figure}[ht]
\centerline{\includegraphics[width=\columnwidth]{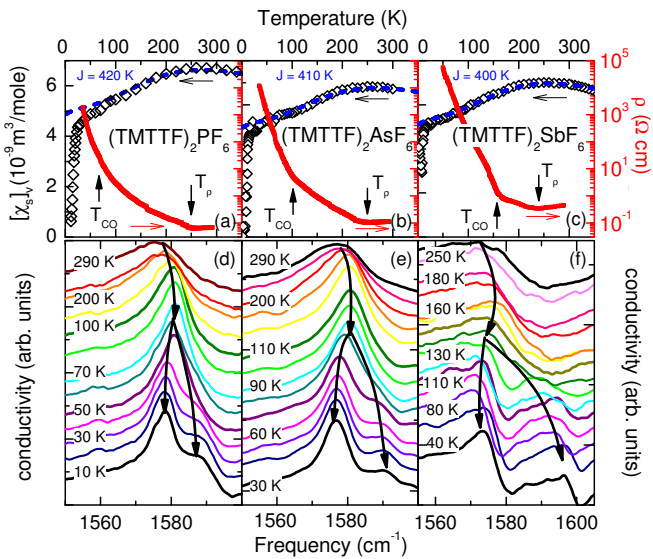}}
\caption{ \label{fig:dc} Electronic, magnetic and
optical characterization of the three compounds (TMTTF)$_2$PF$_6$,
(TMTTF)$_2$AsF$_6$ and (TMTTF)$_2$SbF$_6$. (a)-(c) The open
symbols correspond to the spin susceptibility
$\left(\chi_S\right)_V(T)$ along the $a$ direction evaluated at
constant volume (left axis). The dashed blue lines correspond to
fits using the model \cite{Eggert94} of a Heisenberg chain with
$J$ as indicated. For (TMTTF)$_2$PF$_6$ and (TMTTF)$_2$AsF$_6$,
$(\chi_S)_V(T)$  drops exponentially at $T_{\rm SP}=19$ and 13~K,
respectively, indicating a transition to a nonmagnetic
spin-Peierls ground state. In contrast, (TMTTF)$_2$SbF$_6$ orders
antiferromagnetically at $T_N=8$~K. The
solid red dots show the temperature-dependent dc resistivity
$\rho(T)$ measured along the $a$ axis (right scale). The solid
arrows indicate the charge localization at $T_{\rho}$ and the
charge-order transition $T_{\rm CO}$  \label{fig:chi}
\label{fig:nu}(d)-(f) Temperature development of the asymmetric
vibration of the TMTTF molecule involving the C$=$C bonds obtained
with light polarized perpendicular to the $ab$ plane. For
$T<T_{\rm CO}$ the mode splits due to charge imbalance.}
\end{figure}

In these strongly correlated electron systems, nearest-neighbor
Coulomb repulsion causes CO upon cooling below $T_{\rm CO}$
characterized by a charge gap in the transport properties
\cite{Kohler11}. From the magnetic point of view, these are
$S=1/2$ antiferromagnetic chains with some ordering at low
temperatures \cite{Dumm00,Salameh11}. Surprisingly, by now no
indications of the  CO transition could be observed in the
magnetic properties, such as spin susceptibility (cf.\
Fig.~\ref{fig:dc}). Here we report comprehensive electron spin
resonance (ESR) investigations that reveal a novel magnetic
interaction mechanism caused by charge disproportionation; we
observe first evidence that charge order breaks the
symmetry of the magnetic degree of freedom in these quantum spin
chains.

\section{Results and Analysis}
Single crystals of (TMTTF)$_2X$ with octahedral anions $X$ =
PF$_6$, AsF$_6$, and SbF$_6$, are grown electrochemically. The planar TMTTF molecules stack in a slight
zig-zag fashion normal to the molecular plane; this $a$ axis is
the best-conducting direction. The stacks are arranged next to
each other with some minor interaction in $b$ direction and
separated in the $c$ direction by the anions. Since one electron
is transferred to each anion, the  organic molecules are left with
half a hole, on average. Triggered by the A$_2$B stoichiometry,
the TMTTF molecules form dimers along the $a$-axis;
a fact that enhances the charge localization observed
as a broad minimum in resistivity at
$T_{\rho} \approx 250$~K
\cite{Jerome82,Pouget96,Ducasse96,Dressel03}. In
Fig.~\ref{fig:dc}(a)-(c), $\rho(T)$ is plotted for the three
compounds under consideration \cite{Kohler11}.

\begin{table}
\caption{Transition temperatures for charge localization
$T_{\rho}$, charge-order $T_{\rm CO}$, spin-Peierls transition
$T_{\rm SP}$ and N{\'e}el order $T_N$ of different  (TMTTF)$_2X$
salts. $J$ denotes the antiferromagnetic exchange constant
obtained from susceptibility measurements fitted according to
\cite{Eggert94}, $2\delta$ is the charge imbalance between two
TMTTF molecules (for $T\rightarrow 0$) measured by vibrational
spectroscopy and $\Delta_{\rm CO}$ indicate the zero-temperature
charge gap derived from dc transport along the $a$-direction.
$\phi$ is the angle the ${\bf g}$-tensor rotates around the
molecular axis in the CO state determined at $T\approx T_{\rm
CO}/2$.\label{table}}
\begin{center}
\begin{tabular}{l|cccccccc}
\hline\hline
compound & $~T_{\rho}$ & $~T_{\rm CO}$  & $~T_{\rm SP}$ & $T_N$& $J$ & $2\delta$ &$\Delta_{\rm CO}$ & $\phi$\\
 & (K) & (K) & (K) & (K) &(K)&$(e)$ &(K) &($^{\circ})$\\
\noalign{\smallskip}\hline\noalign{\smallskip}
(TMTTF)$_2$PF$_6$ & 250 & 67 & 19& -~&420 & 0.15 & small & 22\\
(TMTTF)$_2$AsF$_6$& 250 & 102 & 13 &-~&410 & 0.21 & 310 & 22 \\
(TMTTF)$_2$SbF$_6$& 240 & 157 & ~-~ &8 &400 & 0.29 &745 & 32\\
\hline\hline
\end{tabular}
\end{center}
\end{table}

In addition, basically all TMTTF salts develop charge imbalance
for $T <T_{\rm CO}$, resulting in charge-rich ($0.5e + \delta$)
and charge-poor molecules ($0.5e - \delta$) as first proven by
nuclear magnetic resonance \cite{Chow00}. The charge
disproportionation can be best visualized by the splitting of the
intramolecular vibration $\nu_{28}$(B$_{1u})$ plotted in
Fig.~\ref{fig:nu}(d)-(f)  \cite{remark3}. The results are
summarized in Tab.~\ref{table}; for further details, see Ref.~\cite{Yasin12}.

Charge order causes an additional  gap to open in the density of
states at $T_{\rm CO}$, leading to a kink in the electrical
resistivity $\rho(T)$ indicated by arrows in Fig.~\ref{fig:dc}(a)-(c); as
the temperature decreases it evolves in a mean-field fashion to
$\Delta_{\rm CO}(T\rightarrow 0)\approx 745$~K, for the example of
(TMTTF)$_2$SbF$_6$ \cite{Kohler11}.

Although the effect of charge ordering on the spin degree of
freedom has attracted considerable attention from the theoretical
side \cite{Tanaka06}, it is not completely clear how the spin and
orbital magnetic moments behave in the charge-order state and in
the metallic state under the influence of charge fluctuations.
Naively one would expect that a phase transition, which
has such a severe effect on the electric transport, evidences also
in the spin susceptibility, albeit spin-charge separation is an
issue for one-dimensional conductors \cite{Dressel03}. Despite
numerous charge-ordered systems investigated over the years, no
sign of the charge-order was revealed in the magnetic behavior.
This is demonstrated in Figs.~\ref{fig:chi}(a)-(c) which show
$\chi_S(T)$ for various TMTTF salts \cite{Dumm00,Salameh11}. The
obtained coupling constants $J$ are listed in Tab.~\ref{table}.

\begin{figure}
\centerline{\includegraphics[width=0.8\columnwidth]{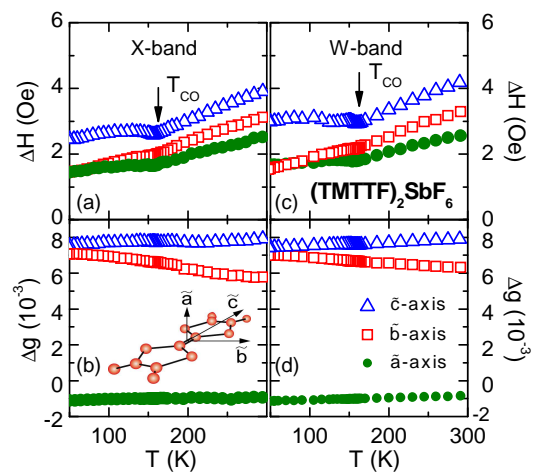}}
\caption{ Temperature dependence of the
linewidth $\Delta H$, and the change in the $g$-value for
(TMTTF)$_2$\-SbF$_6$ measured by X- (left panels) and  W-band (right panels) ESR along three orientations.
The green solid dots correspond to the $\tilde{a}$ axis, the
open red squares indicate the $\tilde{b}$ direction, and the blue
triangles are measured along the $\tilde{c}$ direction. The inset exhibits the TMTTF molecule with its
molecular axes. \label{fig:SBF6}}
\end{figure}
We have performed comprehensive multifrequency ESR investigations
(X-, Q- and W-band) in the temperature range from 4 to 300~K at
different orientations of the magnetic field and by rotating
around the molecular axes (cf. Fig.~\ref{fig:S1}). In
Fig.~\ref{fig:SBF6} the linewidth $\Delta H(T)$ and the $g$-shift
(from the free electron value $g_e = 2.002319$) are plotted as a
function of temperature for the example of (TMTTF)$_2$SbF$_6$. The
$g$-factor is determined by the $\pi$
electronic wavefunction on the TMTTF molecule, the anisotropy
arises from spin-orbit coupling of the conduction electrons
\cite{Coulon04,Furukawa09}; in other words the arrangement of the
${\bf g}$-tensor can be solely deduced from the molecular
symmetry. At room temperature the $g$-factor and linewidth are
maximum when the external magnetic field is oriented along the
molecular axis $\tilde{c}$, while $g$ and $\Delta H$ are smallest
when measured normal to the TMTTF molecules ($\tilde{a}$ axis
which is along the stacks). No indications of the charge-order
transition are observed in the $g$-shift; $\Delta g(T)$ is
basically temperature independent except a small reduction of the
in-plane anisotropy upon cooling. The intriguing  observation is the
clear anomaly in the linewidth at $T_{\rm CO}$ which can also be
seen in \cite{Nakamura03}: the linear decrease in $\Delta H(T)$
becomes weaker indicating an additional interaction
channel to develop for $T<T_{\rm CO}$. Basically the same behavior
at $T_{\rm CO}$ is observed for (TMTTF)$_2$AsF$_6$, while for
(TMTTF)$_2$PF$_6$ the effect is smaller due to stronger coupling
$J$ and  lesser charge imbalance $2\rho$, which coincide with the
much lower transition temperature $T_{\rm CO}$ and almost
vanishing charge gap $\Delta_{\rm CO}$, summarized in
Tab.~\ref{table}.

\begin{figure}
\centerline{\includegraphics[width=\columnwidth]{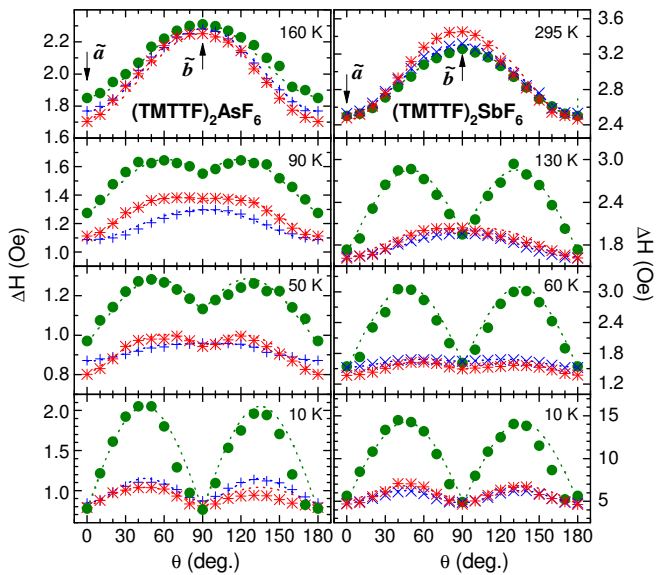}}
\caption{ \label{fig:DeltaH}
 Angular dependence of the ESR linewidth
within the $\tilde{a}\tilde{b}$ plane of (TMTTF)$_2$AsF$_6$  and
(TMTTF)$_2$SbF$_6$ measured at different temperatures above and
below the charge-order transition ($T_{\rm CO} = 102$ and 157~K,
respectively) using X-band (blue crosses), Q-band (red stars), and
W-band (green dots) spectrometers. The lines represent fits of the
linewidth using Eqs.~(\ref{eq:spinphonon}) and (\ref{eq:CO})}
\end{figure}
\begin{figure}
\centerline{\includegraphics[width=0.60\columnwidth]{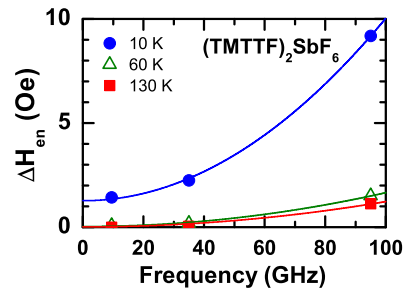} }
\caption{Frequency dependence of the enhanced ESR
linewidth $\Delta H_{\rm en}$ of
(TMTTF)$_2$SbF$_6$ measured in the $\tilde{a}\tilde{b}$ plane
along the diagonal direction ($\theta=45^{\circ}$).
The data are shown at $T=10$~K (blue dots), 60~K (green
triangles) and 130~K (red squares). The lines are fits using
Eq.~(\ref{eq:nu2}).\label{fig:nu2}}
\end{figure}

In order to gain more insight how charge order affects the spins,
we performed detailed angular-dependent measurements of $\Delta H$
and $\Delta g$ at selected temperatures. While the rotational
symmetry of the $\Delta g$ remains unchanged with temperature,
$\Delta H$ exhibits a very unusual behavior as demonstrated in
Fig.~\ref{fig:DeltaH}. For $T>T_{\rm CO}$ the $g$-factor and the
linewidth are dominated by spin-phonon scattering with an angular
dependence \cite{Dumm00}
\begin{equation}
\Delta H_{sp} (\theta) = \left[\Delta H_{sp}^{2}(\tilde{a})\cos^2\theta + \Delta H_{sp}^{2}(\tilde{b})\sin^2\theta\right]^{1/2}
\label{eq:spinphonon}
\end{equation}
in the $\tilde{a}\tilde{b}$ plane, for instance. Most remarkably, the periodicity doubles as the charge-ordered state is entered. In the CO state the
orientational dependence of the linewidth can be fitted by the sum
of Eqs.~(\ref{eq:spinphonon}) and (\ref{eq:CO}):
\begin{equation}
\Delta H_{\rm mod}(\theta) = \Delta H_{\rm mod}(45^\circ) ~|\sin\{2\theta\}| \quad ,
\label{eq:CO}
\end{equation}
where $\Delta H_{\rm mod}$ is the excess linewidth.
The enhancement of the linewidth observed in the diagonal
direction $\Delta H_{\rm en}= \Delta H(45^{\circ}) - \frac{1}{2}
[\Delta H (\tilde{a}) + \Delta H (\tilde{b})]$ becomes more
pronounced as the frequency $\nu$ increases, as
depicted in Fig.~\ref{fig:nu2}. The quadratic
behavior
\begin{equation}
\Delta H_{\rm en} (\nu) =A + B \nu^2 \label{eq:nu2}
\end{equation}
 strongly indicates
anisotropic Zeeman interaction \cite{Bencini90}. The same period
doubling in $\Delta H(\theta)$ can be observed in our X-band
experiments on (TMTTF)$_2$PF$_6$, although the effect is much
weaker.

In the case of staggered fields, antisymmetric
Dzyalo\-shinskii-Moriya interaction also causes a $\nu^2$ frequency dependence of
the linewidth; however, leading to a $T^{-2}$ temperature dependence \cite{Oshikawa99,Herak11}
in contrast to our findings. Note also that in those references
Heisenberg spin chains with significant intra-chain Dzyaloshinskii-
Moriya interaction were considered. In the present case, the
alternating contributions result from interaction between the chains,
for which the iostropic exchange $J^{\prime}$ is already small. Thus the
antisymmetric Dzyaloshinskii-Moriya exchange can be neglected.

>From similar observations on inorganic linear-chain compounds,
 Pilawa derived a contribution of the anisotropic Zeeman interaction to the linewidth \cite{Pilawa97}
\begin {equation}
\Delta H_{\rm AZ} \approx \frac{\mu_BH^2_0}{g_e k_B\, | J^{\prime} |}
\sqrt{\frac{\pi}{8}}~ |\Delta g |^2
\quad , \label{eq:AZ}
\end {equation}
where $J^{\prime}$ is the exchange interaction between two
inequivalent magnetic sites on the neighboring chains; $\Delta g$ is the difference between their ${\bf
g}$-tensors. In our case
the interchain exchange constant is $J^{\prime}=1.1$~K
\cite{Yasuda05}.
With $B_0=33.6$~kOe for the W-band, we obtain
$\Delta g = 7.82\times 10^{-3}$ and $16.78\times 10^{-3}$ for
(TMTTF)$_2$AsF$_6$ and (TMTTF)$_2$SbF$_6$, respectively
\cite{remark1}. The fact that we do not observe two distinct  but
only a single Lorentzian line implies that the hopping rate
between the non-equivalent sites is much larger than the splitting
(strong-coupling limit). The difference in Larmor frequencies in
the W-band $\Delta \nu = (\Delta g/g)\times 95$~GHz corresponds to
only 27~mK, which is much smaller than $J^{\prime}$.

The  magnetically inequivalent sites have to be located on
adjacent stacks; intrachain interaction is so strong that it would
suppress the anisotropic Zeeman effect at these frequencies. From
the fact that the maximum of $\Delta H$ occurs at $\theta
=45^{\circ}$ and that $\Delta g$ is not zero, we conclude that the
${\bf g}$-tensors of the two sites do not coincide with the
principle magnetic axes any more but are rotated around the
$\tilde{c}$ axis by an angle $\phi$ in opposite directions as
depicted in Fig.~\ref{fig:structure}; at temperature well below
$T_{\rm CO}$ we obtain $\phi=\pm 22^{\circ}$ for $X$ = PF$_6$,
AsF$_6$ and $\pm 32^{\circ}$ for (TMTTF)$_2$SbF$_6$, respectively
\cite{remark2}.

\section{Discussion}
\begin{figure}
\centerline{\includegraphics[width=\columnwidth]{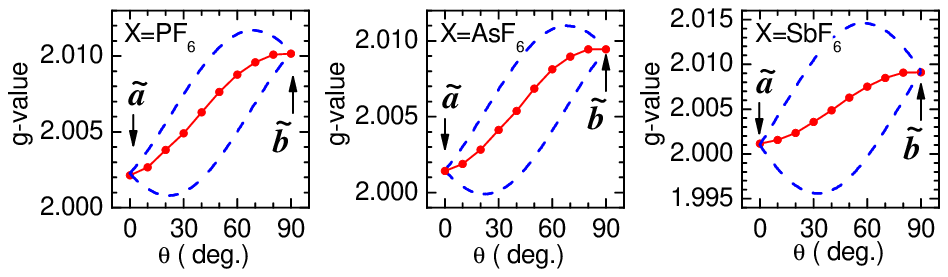}}
\vspace*{2mm}
\centerline{\includegraphics[width=\columnwidth]{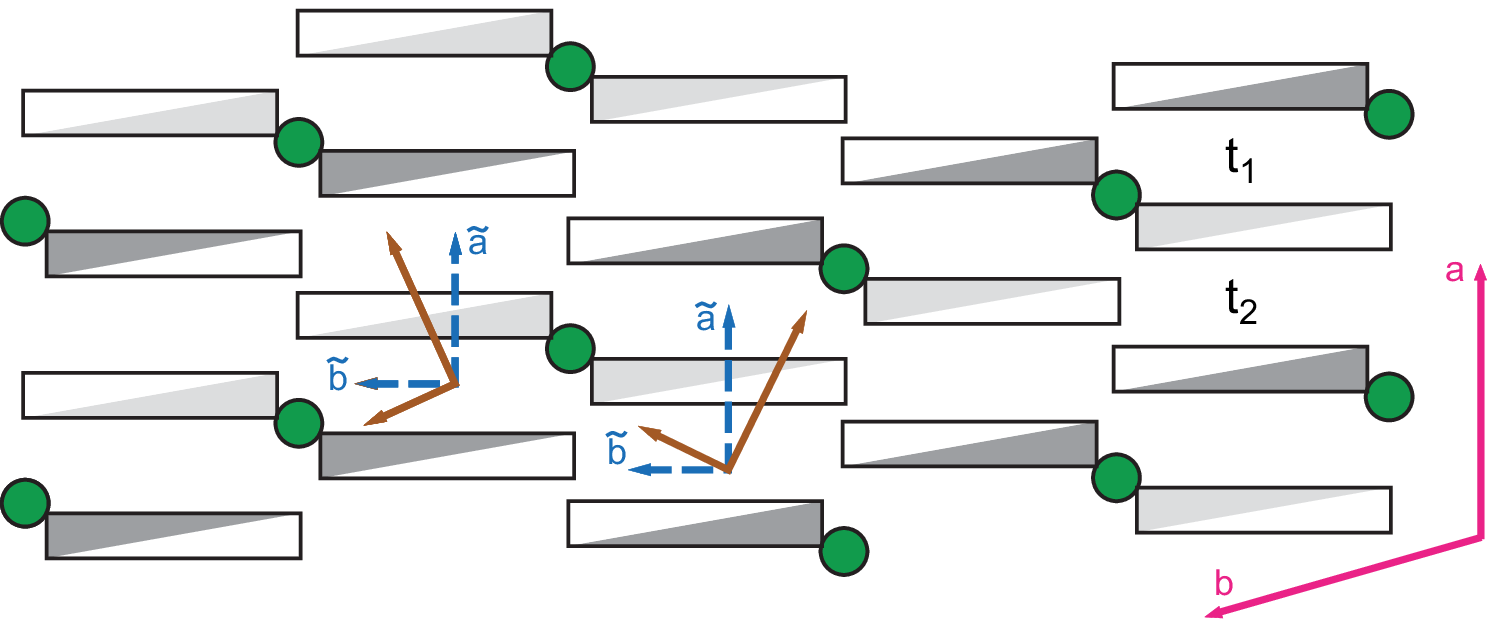} }
\caption{ Angular dependence of the $g$-value of
(TMTTF)$_2$PF$_6$, (TMTTF)$_2$AsF$_6$ and (TMTTF)$_2$SbF$_6$
evaluated at $T=30$, 50 and 60~K, respectively, when a static
magnetic field $B_0$ is rotated in the $\tilde{a}\tilde{b}$ plane.
The dashed lines indicate the calculated variation of the
$g$-values of the two magnetically inequivalent sites. 
\label{fig:g-decompose} A view
along the molecular axis $\tilde{c}$ is sketched in the lower
panel. Below $T_{\rm CO}$ the charge distribution on the dimer
becomes imbalanced (indicated by the different grey shade). The
distinct coupling to the anions (green circles) modifies the
electronic wavefunction causing the $\bf{g}$-tensor to rotate in
opposite directions, as depicted by the brown arrows. The transfer
integrals $t_1$ and $t_2$ differ by about 40\%\ at room
temperature. \label{fig:structure} }
\end{figure}

The crucial question now is, why CO triggers a rotation of the
${\bf g}$-tensors. ESR experiments by Furukawa {\it et al.}
\cite{Furukawa09} revealed that the Coulomb potential of the
anions influences the electronic wavefunctions on the TMTTF
molecules. Depending on whether the adjacent molecules are charge
rich or charge poor, this constitutes non-equivalent charge and
spin distributions as sketched in Fig.~\ref{fig:structure}. The
imbalance of the spin density causes a rotation of the ${\bf g}$
tensor by an angle of $\phi$. Our findings are in accord with the
suggestion by Riera and Poilblanc \cite{Riera01} that charge order
in TMTTF salts is a cooperative effect between the Coulomb
interaction and the electronic coupling of the TMTTF stacks to the
anions \cite{remark4}.

The occurrence of spontaneous di\-pole moments \cite{Monceau01}
and second harmonic generation \cite{Yamamoto11} evidences the loss
of inversion symmetry (with one center located at the anions, for
instance) for $T<T_{\rm CO}$. Our vibrational spectra
(Fig.~\ref{fig:nu}) show the presence of charge-rich and charge
poor molecules, which, however, can be arranged on the dimerized
stacks in two possible ways: $+-~~+-~~+-$ or $-+~~-+~~-+$. This
immediately implies that the rotation of the ${\bf g}$ tensor can
occur in both directions $\pm\phi$. Thus at $T<T_{\rm CO}$ domains
are formed with inequivalent magnetic sites on neighboring stacks
at the boundaries, as depicted in Fig.~\ref{fig:structure}.
Without interaction two distinct ESR signals are expected below
$T_{\rm CO}$, but the strong coupling leads to a significant
broadening, that might get resolved by going to even higher
frequencies.

Structural investigations could not identify any superstructure or
change in symmetry, however, a strong increase in mosaicity at
$T_{\rm CO}$, as shown in Fig.~\ref{fig:S0} \cite{Rose12}. We conclude that the ordering
occurs on a local scale with nanometer-sized CO domains. Similar,
but larger ferroelectric domains of electronic origin have been
directly observed in the two-dimensional organic compound
$\alpha$-(BEDT-\-TTF)$_2$I$_3$ below the CO transition
\cite{Yamamoto10}.

As the averaged ${\bf g}$-tensor does not abruptly change at
$T_{\rm CO}$, the charge ordering can be pictured as follows: For
$T> T_{\rm CO}$ random charge disproportionation exists on the
TMTTF stacks giving rise to an averaged ${g}$-factor  as observed,
but without any anisotropic Zeeman effect, which is suppressed due
to narrowing by large intrachain exchange. Below $T_{\rm CO}$, the
charge imbalance becomes cooperative within each stack;
there is some weak interaction of adjacent chains.
At the boundary of two domains defined by
the direction of charge and spin distortion, translational
symmetry is lost. This gives rise to two inequivalent chains that
yield the same averaged ${\bf g}$-tensor along the $\tilde{a}$ and
$\tilde{b}$ axes as above $T_{\rm CO}$. In the diagonal
directions, however, the anisotropic Zeeman effect now becomes
visible in our ESR data because the signal is narrowed by the much weaker
interchain exchange.

\section{Conclusions}
In summary, the charge order observed in the linear-chain
compounds (TMTTF)$_2X$ by the splitting of the molecular
vibrations opens a gap in the charge excitations. It also affects
the magnetic behavior via mutual interaction with the anions. For
$T<T_{\rm CO}$ the inversion symmetry is broken; charge
disproportionation leads to a different electronic distortion on
the organic molecules that eventually rotates the ${\bf
g}$-tensors, with two directions $\pm \phi$ possible. From our
multifrequency ESR measurements we conclude that in the
charge-ordered regime two inequivalent magnetic TMTTF chains
coexist, which produce a doubling in the angular periodicity of
the linewidth as well as the characteristic quadratic frequency
dependence. Thus, our experiments show that charge order not
only causes ferroelectricity but can also
break the symmetry of the magnetic degree of freedom in quantum
spin chains.

\section{Acknowledgments}
We thank S. Brown, A. J{\'a}nossy, J.-P. Pouget  and  E. Rose for useful
discussions. The crystals were grown by G. Untereiner; H.J.
K\"ummerer, H.-A. Krug von Nidda, F. Lissner and M. Ozerov
assisted during some experiments. The work is supported by the
Deutsche Forschungsgemeinschaft (DFG). T.K. acknowledges a
fellowship of the Carl-Zeiss-Stiftung.

\appendix

\section{Appendix A: Structure}

The Fabre salts (TMTTF)$_2X$ are charge transfer salts consisting
of stacks of the planar organic molecules TMTTF (which stands for
tetra\-methyl\-tetra\-thia\-fulvalene) along the $a$-axis that are
separated in $c$ direction by monovalent anions $X$. The overlap
of the $\pi$ orbitals in $a$ direction makes them model systems of
one-dimensional metals. In $b$-direction the distance of the
stacks  is comparable to the van der Waals radii. Basically no
coupling between the molecular stacks exists in the $c$ direction.
In the case of the Bechgaard salts the TMTSF molecule contains
selenium instead of sulphur with more extended orbitals, leading
to better metallic conduction and even superconductivity
\cite{Jerome82,Jerome94,Ishiguro98,Brazovskii08}.

\begin{figure*}
\centerline{\includegraphics[width=0.35\textwidth]{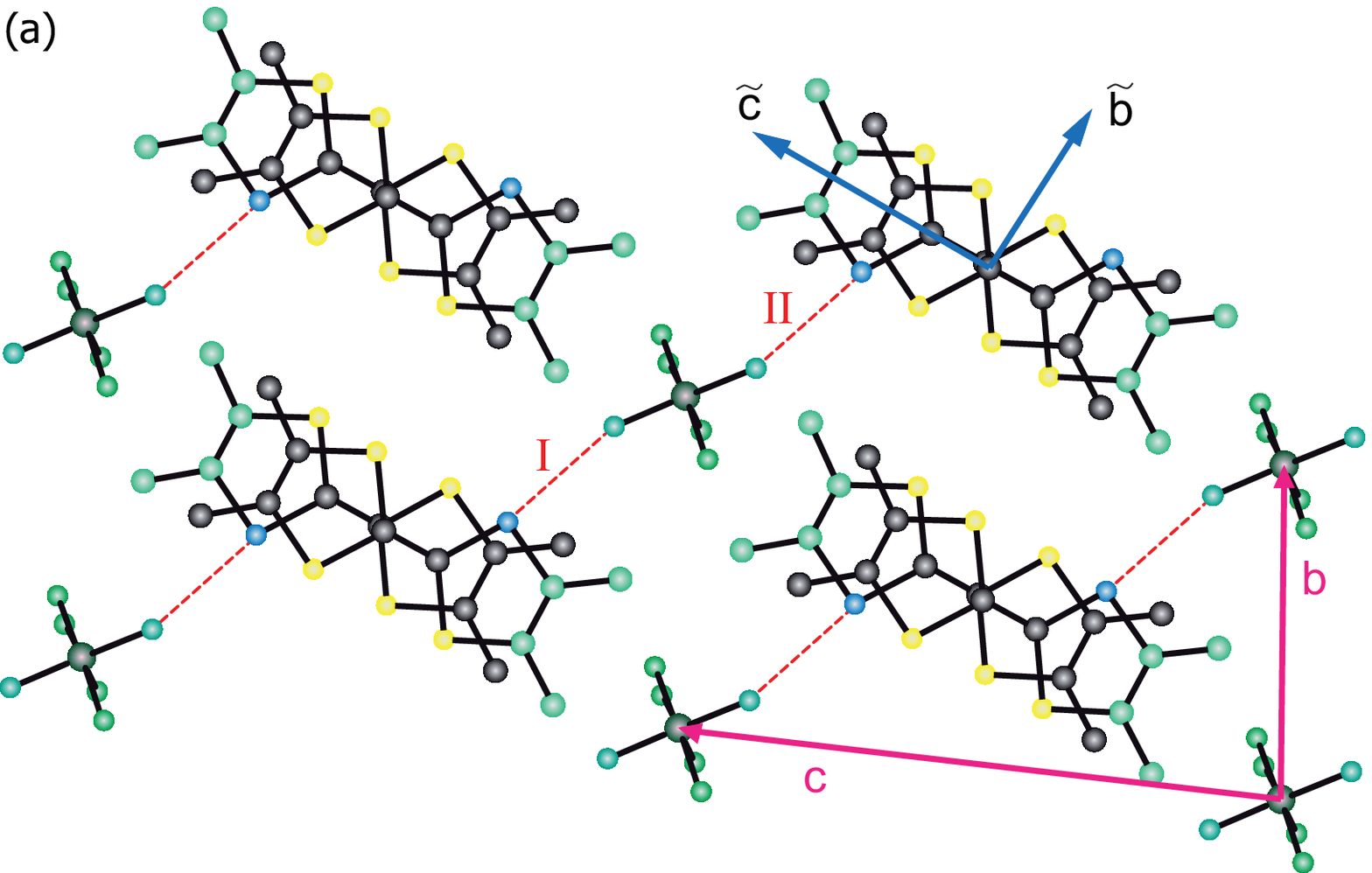}
~~~\includegraphics[width=0.3\textwidth]{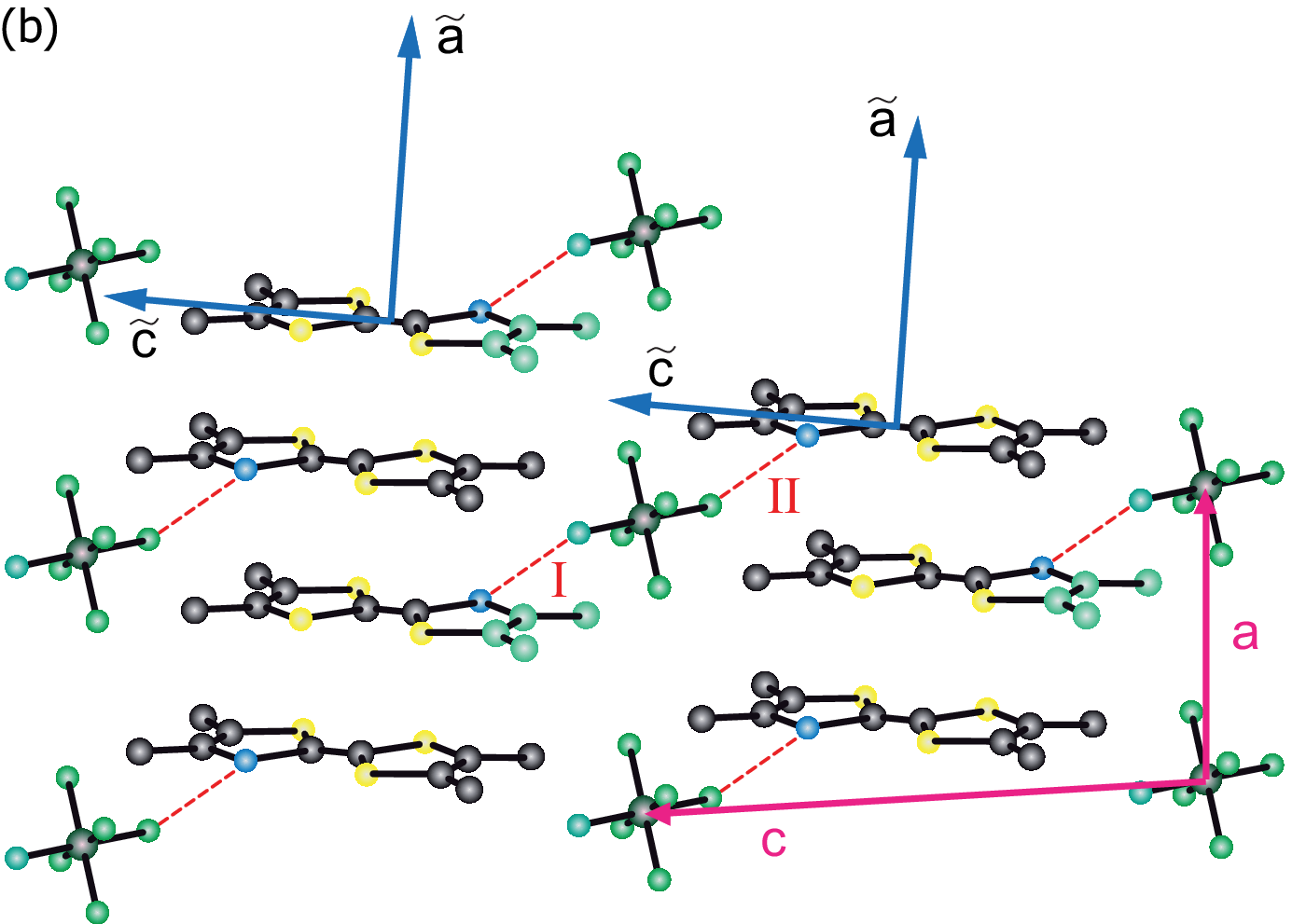}
~~~\includegraphics[width=0.3\textwidth]{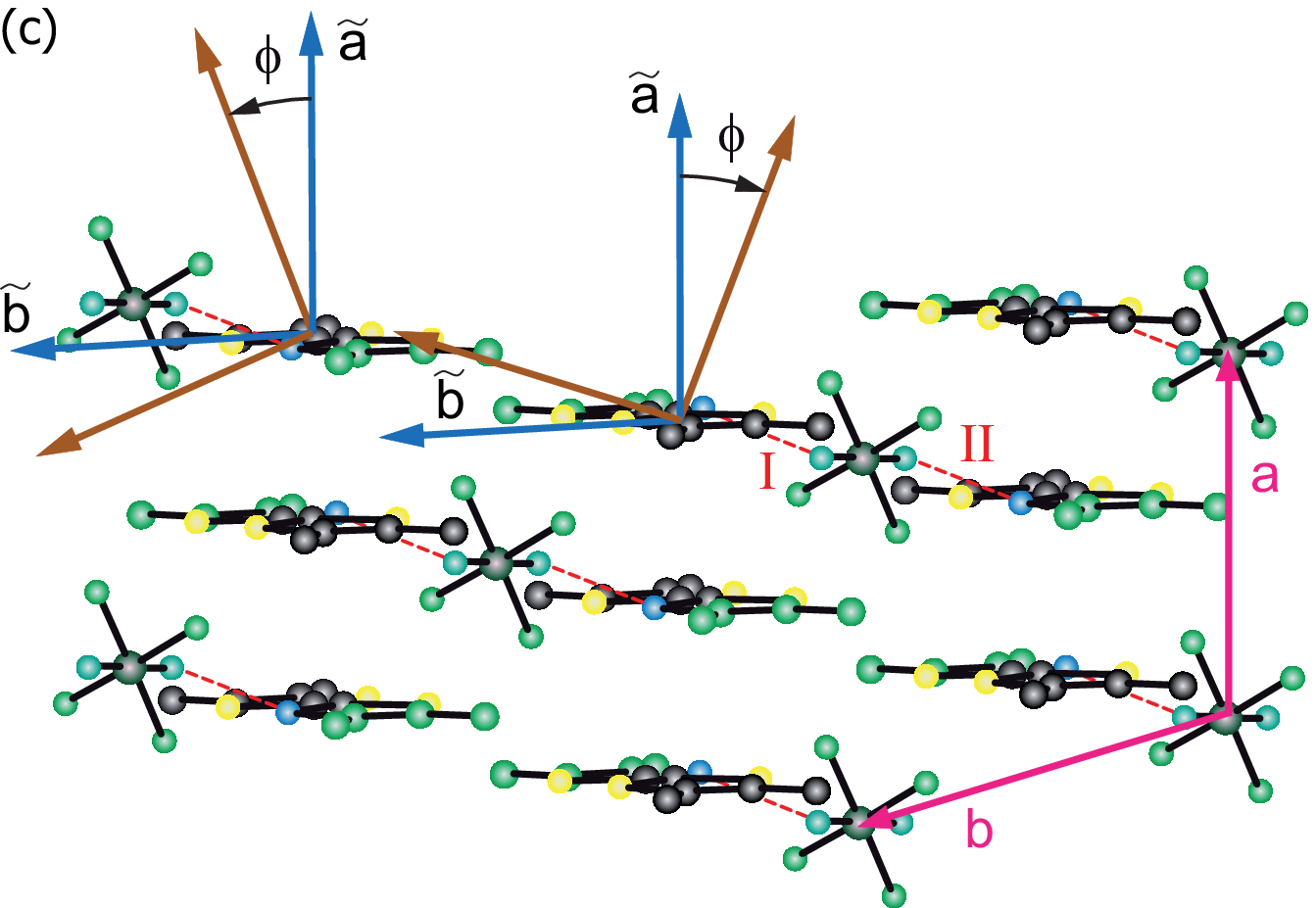} }
\caption{ Crystal structure of (TMTTF)$_2$SbF$_6$,
as an example, viewed (a)~along the stacking direction $a$, (b)
along the $b$ axis and (c)~along the $c$~direction. The strongest
coupling between the octahedral anions and the sulfur atoms are
indicated by the red dashed lines I and II which are related by
inversion symmetry for $T>T_{\rm CO}$. In the charge-ordered
state, not only the TMTTF molecules develop a charge
disproportionation with alternating charge-rich and charge-poor
molecules along the stacks, also the octahedral anions are
slightly distorted. The inversion symmetry is lifted and contact I
becomes stronger than II, for instance. Two kind of domains are
formed separated by domain walls preferentially in the $ab$ or
$ac$ planes. Also shown is the configuration between the TMTTF
molecular structure and the principal axes of the ${\bf
g}$-tensor. Note there exists only one electron per dimer. At
elevated temperatures the ${\bf g}$-tensor (blue arrows indicate
the principal magnetic directions $\tilde{a}$, $\tilde{b}$ and
$\tilde{c}$) is determined by the molecular structure. The
$\tilde{a}$ axis of the $g$-tensor is normal to the molecular
plane and along the stacking direction, while the $\tilde{c}$ axis
points along to the longest molecule extension. For $T<T_{\rm CO}$
the principal axes of the $g$-tensor rotate around the
$\tilde{c}$-axis by an angle $\phi$ in opposite directions,
resulting in the brown arrows. \label{fig:S1} }
\end{figure*}
All compounds of the TMTTF family are isostructural (triclinic
P$\bar{1}$ space group, $C_i$ symmetry with $Z=1$) with the TMTTF stacked
normal to the molecular plane in $a$ direction, slightly displaced
in $c$ direction to optimize the orbital overlap of the selenium
atoms along $a$. In spite of the $A_2B$ stoichiometry and the
corresponding dimerization, the separation of the TMTTF molecules
along the stacks only differs by approximately 3\%\ at room temperature
and even less at low temperature \cite{Laversanne84,Kohler11},
implying almost equal overlap integrals along the $a$ axis.
Due to the triclinic symmetry, $b^{\prime}$ commonly denotes
the projection of the
$b$ axis perpendicular to $a$, and $c^*$ is normal to the $ab$
plane. Here we introduce a Cartesian coordinate system linked to
the TMTTF molecule, with $\tilde{a}$ points normal to the
molecular plane, $\tilde{c}$ is the extended molecular axis, and
$\tilde{b}$ is perpendicular to both (cf.\ inset of Fig.~\ref{fig:SBF6}). 
In a good approximation $\tilde{a}$ is parallel to the
stacking axis $a$, and $\tilde{c}$ is almost $c + b$, as depicted in
Fig.~\ref{fig:S1}.

\begin{figure}
\centerline{\includegraphics[width=0.5\columnwidth]{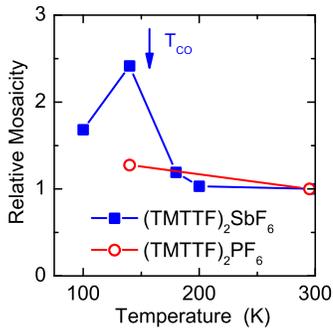}}
\caption{ Temperature dependence of the normalized mosaicity of
(TMTTF)$_2$SbF$_6$ and (TMTTF)$_2$PF$_6$ determined from the width of the x-ray
spots  (after Ref.~\onlinecite{Rose12}). When the temperature drops below 200~K, the mosaicity of (TMTTF)$_2$SbF$_6$ increases dramatically approaching a maximum at the charge order temperature $T_{\rm CO}=157$~K. For comparison the behavior of (TMTTF)$_2$PF$_6$ is also presented where only a small increase is observed. Note, in the case of (TMTTF)$_2$PF$_6$
the phase transition happens only at $T_{\rm CO}=67$~K. \label{fig:S0}}
\end{figure}

X-ray investigations of (TMTTF)$_2$SbF$_6$ single crystals
performed at different temperatures down to $T=80$~K rule out a
structural transition at the charge-order temperature
\cite{Rose12}, confirming previous attempts by C. Coulon and
collaborator \cite{Laversanne84,Coulon85,Granier91}. Structural
data do not unambiguously reveal the loss of inversion symmetry.
Their analysis, however, yields a marginal better fit with no
symmetry restrictions; but this also holds (to a minor degree) at
room temperature. Charge order does not constitute a
superstructure; nevertheless, the spots broaden significantly upon
cooling with a distinct change at $T_{\rm CO}$. This is a measure
of the mosaicity of the sample: in Fig.~\ref{fig:S0} the
temperature dependence of the mosaicity is plotted normalized to
the room-temperature value. We interpret this as the development
of domains with the SbF$_6^-$ anions slightly distorted in one or
the other direction. The distortion, but also the domain size
changes with temperature leading to a maximum around $T_{\rm CO}$.
Typical for an effect of disorder, the phenomenon depends on
cooling rate and slightly vary from crystal to crystal.

\section{Appendix B: Experimental Details}
Single crystals of (TMTTF)$_2X$ with $X$ = PF$_6$, AsF$_6$, and
SbF$_6$  were grown by electrochemical methods in an H-type glass
cell at room temperature \cite{Montgomery94}. A constant voltage
of 1.5~V was applied between platinum electrodes with an area of
approximately 1~cm$^2$. The current through the solution was
between 1 and 2~$\mu$A. After several weeks we were able to
harvest needle-shaped single crystals of several millimeters in
length and less than a  millimeter in width.

The electrical transport was characterized by dc resistivity
measured as a function of temperature in all three directions
\cite{Kohler11}. We evaporated small  gold contacts onto the
natural crystal surface and attached thin gold wires by carbon
paste. Along the long $a$ axis of the crystals and also for the
$b$ direction four-point measurements could be performed, while
for the $c$ direction two contacts were applied on opposite sides
of the crystal. The samples were attached to a sapphire plate in
order to ensure good thermal contact and slowly cooled down to
helium temperatures.

Measurements of the magnetic susceptibility were performed using a
Quantum Design SQUID (superconducting quantum interference device)
magnetometer between 1.8 and 380~K. A large number of single
crystals were glued parallel to each others inside a plastic straw
using vacuum grease, the magnetic field was applied parallel to
the $a$ axis. The background signal of the sample holder and the
vacuum grease was measured separately and subtracted in order to
obtain the intrinsic magnetization of the sample. To estimate the
spin susceptibility, we subtracted a temperature independent
diamagnetic contribution of the core electrons $\chi_{\rm dia} =
-5.3\times 10^{-9} {\rm m}^{3}$/mole from the original data of all
specimen under consideration.

The infrared reflectivity spectra polarized along  the $c$
direction were recorded by utilizing a Bruker IFS 66v/s and a
Bruker Vertex 80 Fourier transform spectrometer extended with a IR
microscope HYPERION which is purged with nitrogen to suppress any
disturbing influence of the water absorption bands. The
reflectivity spectra were measured in a frequency range from
500~cm$^{-1}$ until 8000~cm${-1}$ at temperatures from 10~K up to
290~K. Here we focus on the perpendicular direction, i.e.\
$E\parallel c^*$. To receive the real optical conductivity, we
performed a Kramers-Kronig analysis with a constant extrapolation
for the low-frequency range and a $\omega^{-2}$ and $\omega^{-4}$
decrease for the high-frequency part. With a linear interpolation
between the mode frequency in fully neutral (1628~cm$^{-1}$) and
fully ionized (1548~cm$^{-1}$) TMTTF molecules \cite{Meneghetti84}
we can quantitatively determine the charge imbalance [$2\delta =
\Delta \nu /(80~{\rm cm}^{-1}/e)$], and the molecular ionicity.

The electron spin resonance (ESR) spectra were carried out in a
continuous wave X-band spectrometer Bruker ESP 300 at 9.5 GHz and
a W-band spectrometer Bruker Elexsys 680 at 95 GHz at Stuttgart
University, and a Q-band spectrometer Bruker Elexsys 500 at 34 GHz
at Augsburg University. Complementary X-band measurements were performed
at the Hochfeld-Magnetlabor Dresden. The temperature dependence of the ESR
properties was measured  down to $T=4$~K by utilizing
continuous-flow helium cryostats \cite{Dumm00,Salameh11}. The
typical size of the single crystals used in the X-band ESR
measurements was $2\times 0.5 \times 0.1~{\rm  mm}^{3}$, while
very small samples of less than $0.4\times 0.4 \times 0.05~{\rm
mm}^{3}$ were used in the Q- and W-band ESR measurements in order
to avoid any sample-size effect that  can lead to a broadening of
the linewidth \cite{Yasin12}.

\section{Appendix C: Magnetic Characterization}
The large thermal expansion of the organic compounds has strong
effects on the temperature dependence of the spin susceptibility.
To compare the experimental results (usually obtained at $p = {\rm
const}$) with the theoretical predictions (in general calculated
for $V={\rm const}$), the spin susceptibility at constant pressure
$(\chi_s)_p$ has to be transformed in the spin susceptibility at
constant volume $(\chi_s)_V$. In the case of (TMTSF)$_2$PF$_6$ the
temperature dependence of $(\chi_s)_V$ was estimated by Wzietek
{\it et al.} \cite{Wzietek93} from NMR and x-ray measurements
under pressure. We assumed that the substitution of sulfur for
selenium and the exchange of the inorganic anions has no
considerable influence on the thermal expansion, thus we took the
ratio $(\chi_s)_V /(\chi_s)_p$ for different temperatures to
rescale our susceptibility data of (TMTTF)$_2X$. At high
temperatures the spin susceptibility at constant volume of
(TMTTF)$_2X$ resembles the well-known behavior of a spin-1/2
Heisenberg chain with antiferromagnetic coupling as depicted in
Fig.~\ref{fig:dc}  by the dashed blue lines. The
thermodynamic and magnetic properties of such a system were
studied by Bonner and Fisher \cite{Bonner64}. The advanced model
of Eggert, Affleck, and Takahashi \cite{Eggert94} using the Bethe
ansatz differs appreciably only at low temperatures ($T<0.2 J$).
For $T>100$~K the ESR intensity at constant volume can be modeled
using $J\approx 420$, 410, and 400~K for (TMTTF)$_2$PF$_6$,
(TMTTF)$_2$AsF$_6$ and (TMTTF)$_2$SbF$_6$, respectively.

\begin{figure}
\centerline{\includegraphics[width=0.6\columnwidth]{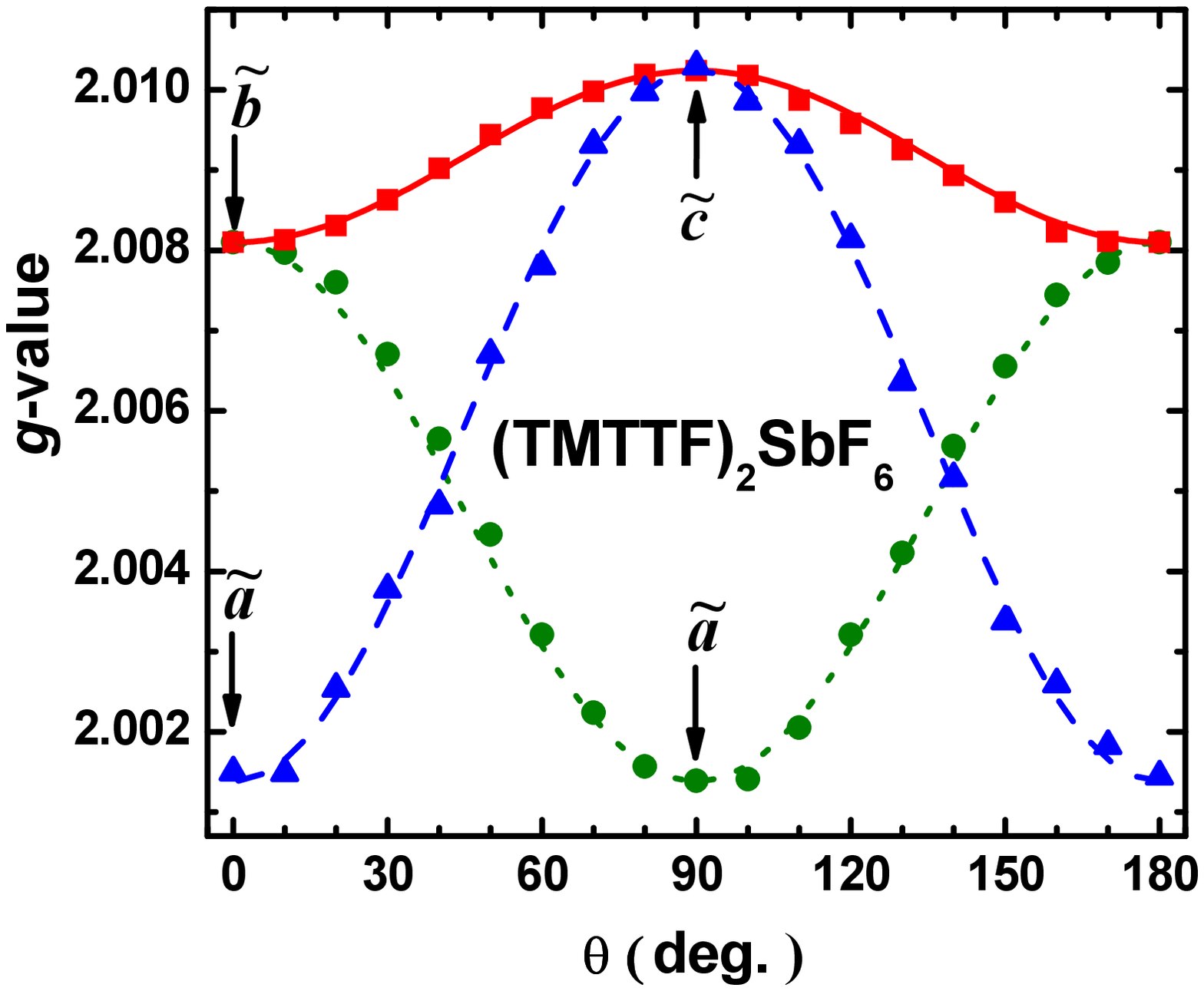}}
\caption{  Angular dependence of the $g$-value
of (TMTTF)$_2$\-SbF$_6$ at room temperature measured at X-band
frequency when the static magnetic field $B_0$ is rotated within
 the $\tilde{a}\tilde{b}$ plane (green circles), the
$\tilde{b}\tilde{c}$ plane (red squares), and the $\tilde{a}\tilde{c}$ plane
(blue triangles). The least-square fits by 
$g(\theta) = \left( g_{\rm min}^{2}\cos^2\theta + g_{\rm
max}^{2}\sin^2\theta\right)^{1/2}$ are
shown by the corresponding lines.
\label{fig:g-angle}}
\end{figure}
\begin{figure*}
\centerline{\includegraphics[width=0.3\textwidth]{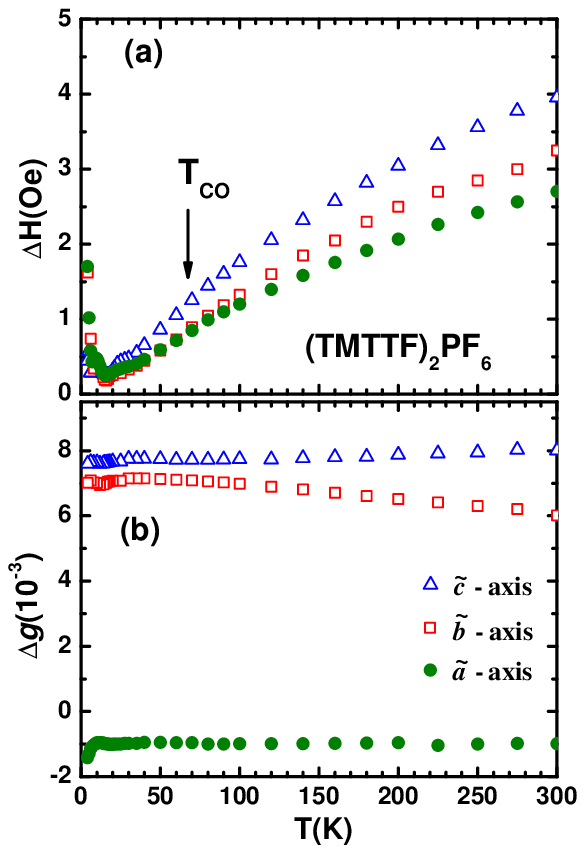}
\includegraphics[width=0.3\textwidth]{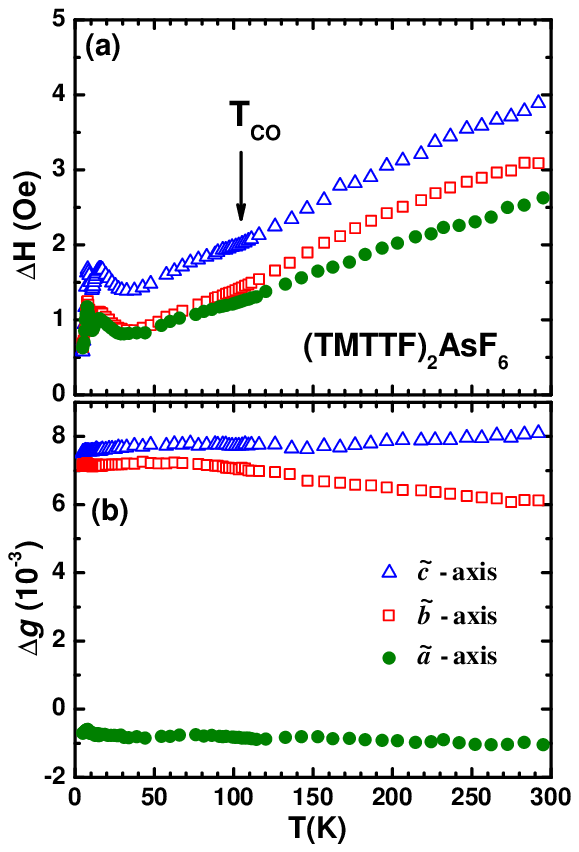}
\includegraphics[width=0.3\textwidth]{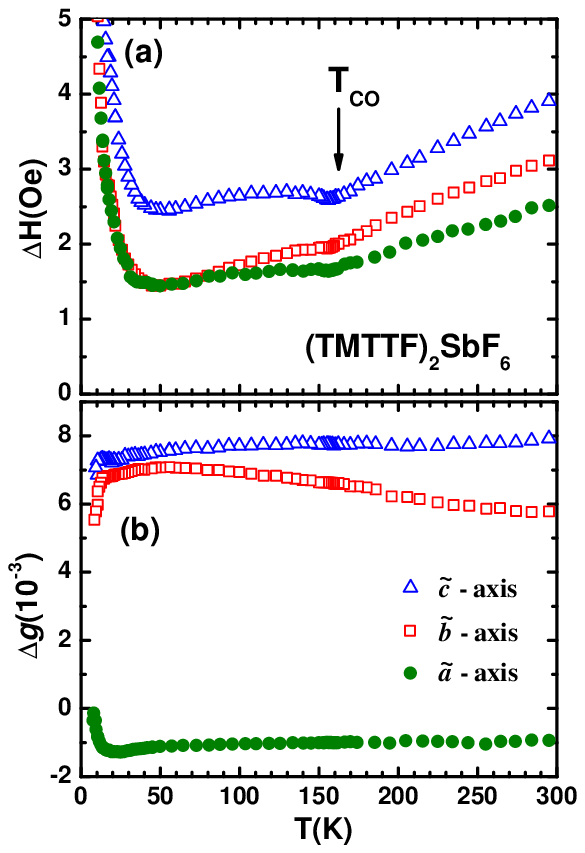}}
\caption{(Color online) Temperature dependence of (a) the ESR
linewidth $\Delta H$, and (b) the $g$-shift $\Delta g(T) = g(T) -
2.002319$ for (TMTTF)$_2$PF$_6$ (left panel), (TMTTF)$_2$AsF$_6$
(middle panel) and (TMTTF)$_2$SbF$_6$ (right panel) measured by
X-band spectroscopy along the three directions $\tilde{a}$ (green
dots), $\tilde{b}$ (open red squares), and $\tilde{c}$ (blue
triangles).\label{fig:S2}}
\end{figure*}

The temperature dependence of the ESR results along the three
crystal axes of (TMTTF)$_2$PF$_6$, (TMTTF)$_2$AsF$_6$ and
(TMTTF)$_2$SbF$_6$ are shown in Fig.~\ref{fig:S2}. The $g$-shift
$\Delta g(T) = g(T) - 2.002319$ for all three crystal axes is very
small. At elevated temperatures, the $g$ values show no
significant temperature dependence but a distinct anisotropy with
a negative value of $\Delta g$ along the chain direction
$\tilde{a}$ and positive values perpendicular to the chain
direction, as seen in Fig.~\ref{fig:g-angle}.
This behavior is found in all TMTTF and TMTSF salts as
presented in Refs.~\cite{Dumm00,Salameh11}. In all cases the small
anisotropy of $\Delta g$ between the $\tilde{b}$ and the
$\tilde{c}$ axes increases with temperature. The $g$ value is
largest for $B_0\parallel \tilde{c}$ and closest to the
free-electron value parallel to the stacks. As the temperature is
lowered, the linewidth $\Delta H$ decreases almost linearly with T
in all directions. In the $\tilde{c}$ direction $\Delta H(T)$ is
slightly larger; the most narrow lines are always observed along
the chains.

At low temperatures (TMTTF)$_2$PF$_6$ and (TMTTF)$_2$AsF$_6$
undergo spin-Peierls transitions at $T_{\rm SP}=19$ and 13~K,
respectively. The formation of states with singlet-paired holes
due to the tetramerization causes a decreasing susceptibility with
an exponential drop of the intensity down to lowest temperatures,
but it does not completely vanish (Fig.~\ref{fig:chi}). This
phase transition also leads to a broadening of the ESR linewidth
and a small change in the $g$ values \cite{Dumm00}.
(TMTTF)$_2$SbF$_6$, on the other hand, orders
antiferromagnetically at $T_N=8$~K, leading to a decrease in
$\chi_s(T)$ with decreasing temperature before it eventually turns
up again.
 This phase transition is accompanied by a
decrease in the $g$-shift for the $\tilde{b}$ and $\tilde{c}$
axes; along the $\tilde{a}$ axis  the $g$ factor becomes larger
below $T\approx 15$~K. The low-temperature ordered states are out
of the scope of the present paper and discussed elsewhere
\cite{Dumm00,Dumm00a}

>From the fact that the maximum of $\Delta H$ occurs at the
diagonal direction  $\theta =45^{\circ}$ of the $ab$ plane and
that the $g$-values of neighboring molecules differ by $\Delta g$
we conclude that the ${\bf g}$-tensors of the spins corresponding
to the two different magnetic sites do not coincide with the
principle magnetic axes. They are rotated around the $\tilde{c}$
axis by $\phi$ in opposite directions as depicted in
Fig.~\ref{fig:S1}. As shown in Fig.~\ref{fig:g-decompose}, the two
different sites contribute to the observed $g$-value according to
\begin{eqnarray}
g_1 &=& \left(g^2_{\rm max}\cos^2\{\theta-\phi\}+g^2_{\rm
min}\sin^2\{\theta-\phi\}\right)^{1/2}
\\
g_2 &=& \left(g^2_{\rm max}\cos^2\{\theta+\phi\}+g^2_{\rm
min}\sin^2\{\theta+\phi\}\right)^{1/2}
\end{eqnarray}
where $g_1$ and $g_2$ correspond to the ${\bf g}$-tensor of type 1
and type 2 spins, respectively. The principal axes of the two
kinds of spins span the angle $\phi$. $g_{\rm max}$ and $g_{\rm
min}$ is the maximum and the minimum of the $g$-values for each
kind of spins at the angle $\phi$, respectively. For symmetry
reasons the $g$ values have to be identical $g_1=g_2$ at
$\theta=0^{\circ}$ and $\theta=90^{\circ}$, while at
$\theta=45^{\circ}$ the difference of the $g$ values is obtained
from the anisotropic Zeeman effect. We obtain for
(TMTTF)$_2$PF$_6$ $g_{\rm max}=2.0117$, $g_{\rm min}=2.0008$, and
$\phi=22^{\circ}$; for (TMTTF)$_2$AsF$_6$ $g_{\rm max}=2.0071$,
$g_{\rm min}=1.9999$, and $\phi=22^{\circ}$; and $g_{\rm
max}=2.0143$, $g_{\rm min}=1.9956$, and $\phi=32^{\circ}$, for the
example of (TMTTF)$_2$SbF$_6$.

\begin{figure*}
\centerline{\includegraphics[width=0.307\textwidth]{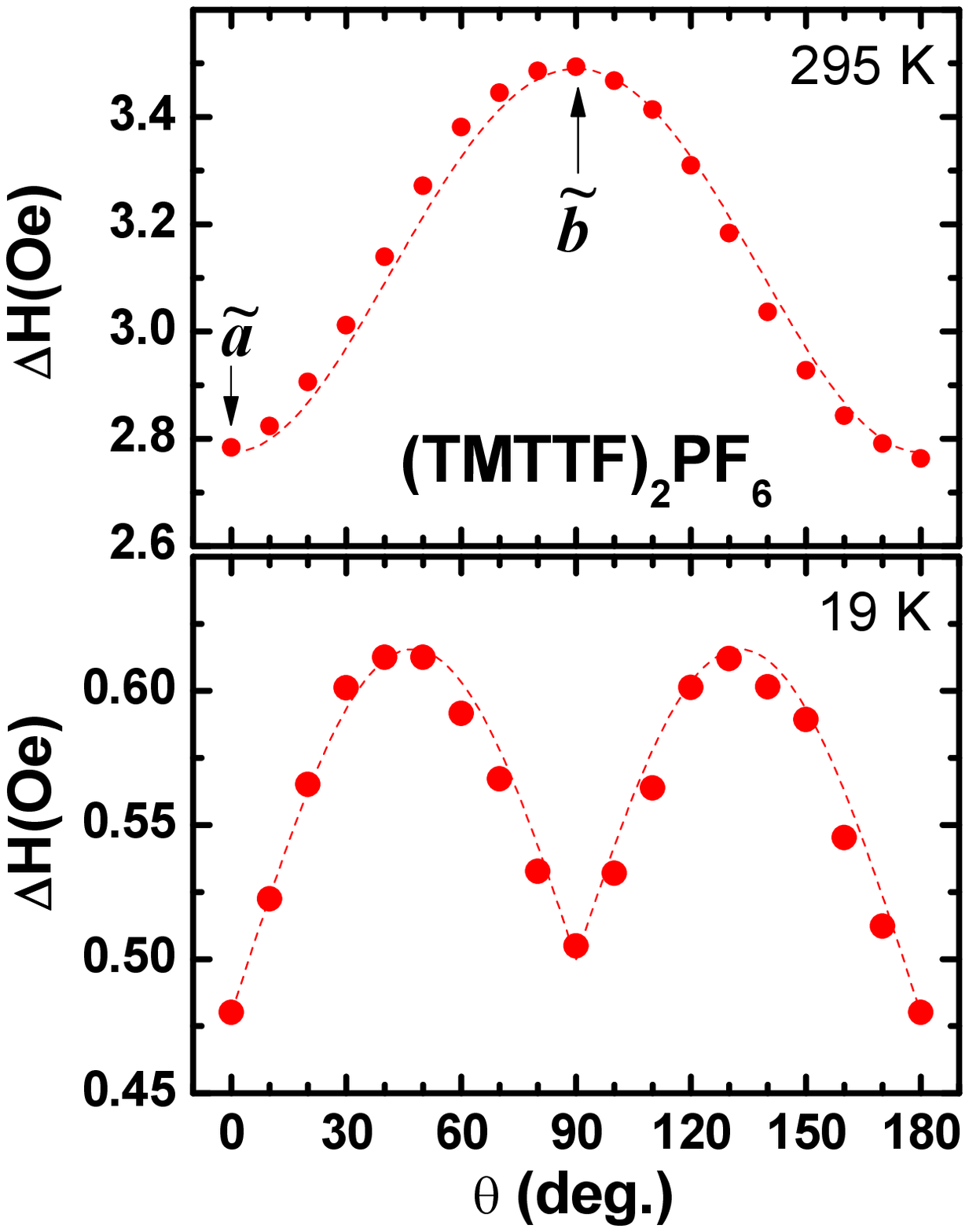}
\includegraphics[width=0.3\textwidth]{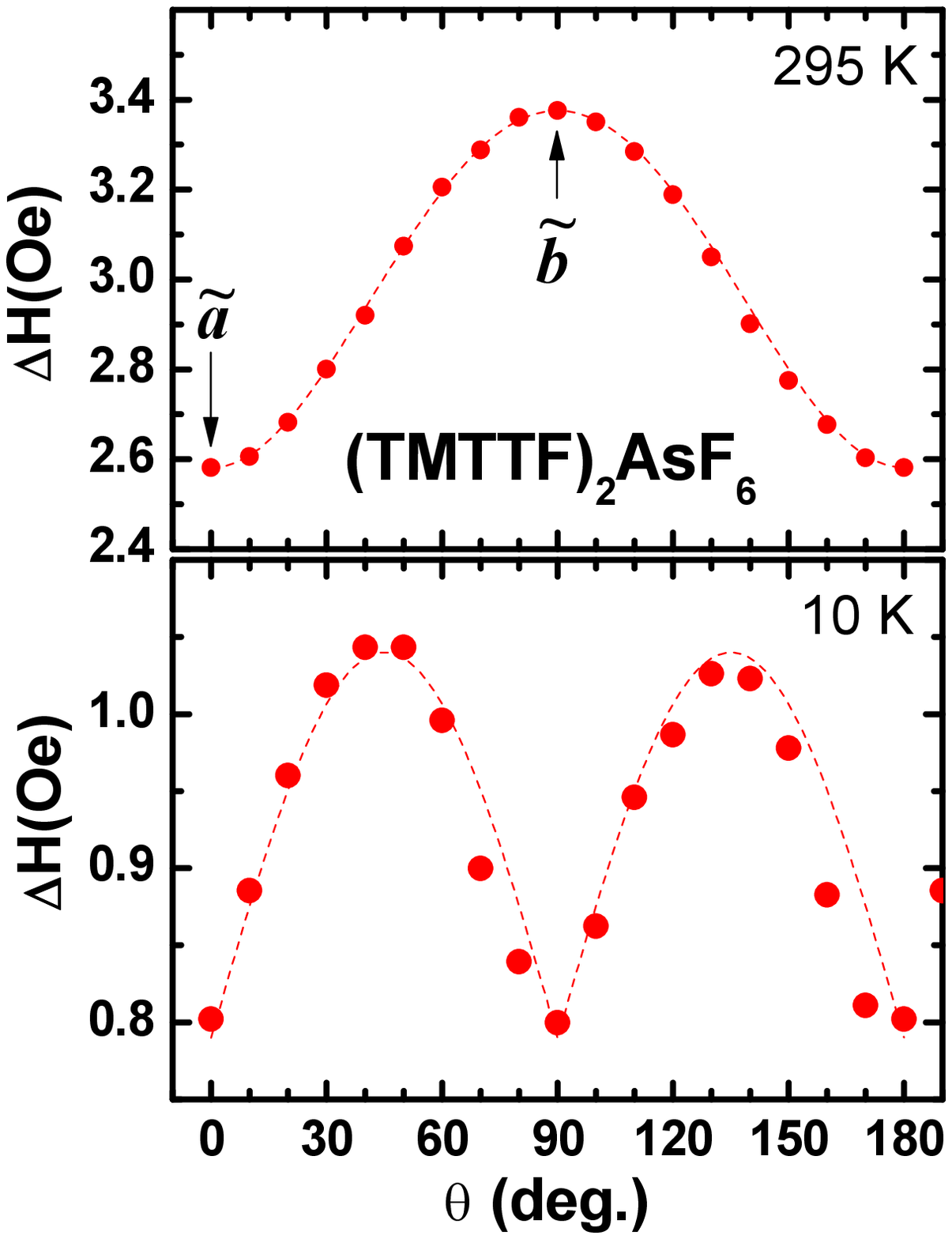}
\includegraphics[width=0.3\textwidth]{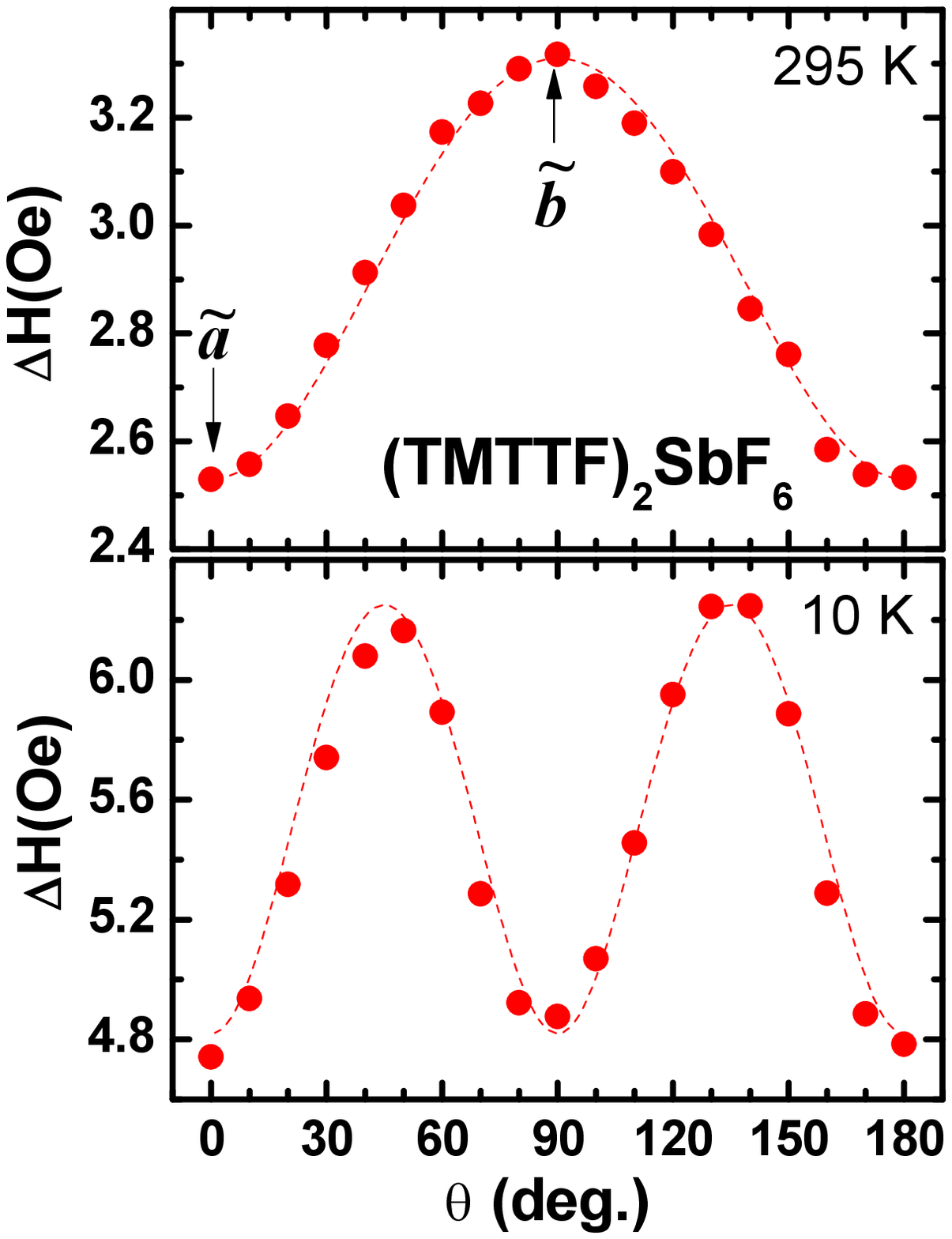}}
\caption{(Color online) Angular dependence of the X-band ESR
linewidth within the $\tilde{a}\tilde{b}$ plane of
(TMTTF)$_2$PF$_6$ (left panel), (TMTTF)$_2$AsF$_6$ (middle panel)
and (TMTTF)$_2$SbF$_6$ (right panel) measured at (a) room
temperature and (b) low temperature, as indicated. The solid lines
are obtained from fits of the data by Eqs.~(1) and (2). Note the
different vertical scale, indicating the observed effect is about
ten times as big in (TMTTF)$_2$SbF$_6$ compared to
(TMTTF)$_2$PF$_6$. \label{fig:S3}}
\end{figure*}
In Fig.~\ref{fig:S3} the linewidth of (TM\-TTF)$_2$\-PF$_6$.
(TM\-TTF)$_2$\-AsF$_6$ and (TM\-TTF)$_2$\-SbF$_6$ is plotted as a
function of angle when the magnetic field is rotated around the
$\tilde{c}$ axis. While at room temperature (upper frames) the
angular dependence follows Eq.~(1), $\Delta H$ strongly increases
in the diagonal direction ($\phi=\pm 45^{\circ}$) as $T<T_{\rm
CO}$ (lower frames). The effect becomes stronger when going from
(TMTTF)$_2$PF$_6$ to (TMTTF)$_2$AsF$_6$ and to (TMTTF)$_2$SbF$_6$,
corresponding to the increase in charge disproportionation and
transition temperature $T_{\rm CO}$.


\begin{thebibliography}{99}
\bibitem{remark1}In the case of (TMTTF)$_2$PF$_6$ we performed
only X-band measurements which in general yield results similar to
the Q-band; we estimate $\Delta g = 7.45\times10^{-3}$.
\end{thebibliography}
\end{document}